LETTER TO THE EDITOR

## Comment on "Heisenberg in Poland" by Jeremy Bernstein
## [Am.J.Phys. 72(3), 300-304 (2004)]

### Klaus Gottstein

Having worked under Werner Heisenberg for twenty years (1950-1970), first as a graduate student and in later years as head of the experimental division of the Max Planck Institute for Physics of which Heisenberg was the director, and having occasionally discussed with Heisenberg some of his experiences during the war, I read with interest the article "Heisenberg in Poland" by Jeremy Bernstein which was published recently in the American Journal of Physics (**72**, 300 (2004)).

My impression is that Dr. Bernstein's paper is not impartial. Why doesn't Bernstein believe Mrs. Heisenberg who should know best the motivations of her husband for his lectures in occupied and neutral countries during the war? Elisabeth Heisenberg's explanation that Heisenberg wanted not to lose contact with his harried colleagues and to demonstrate that a different, better Germany still existed than Nazi Germany sounds very credible. Bernstein does not offer an alternative explanation. He concedes that Heisenberg was not a Nazi, not a member of the party nor of any of its affiliates, and that there were many Jews among his friends, his teachers, and his students. Bernstein mentions that Heisenberg made a special effort to express his regrets to Born, his Jewish teacher, for having received an undivided Nobel Prize, a prize that Heisenberg would have gladly shared with Born. What other explanation is there for Heisenberg's travels during the war than that given by Elisabeth Heisenberg?

I think it is very unfair to hold Heisenberg responsible for the fact that under Nazi policy Polish physicists were not allowed to attend Heisenberg's lecture at the Institut für Deutsche Ostarbeit in Cracow. None of the Cracow physicists (neither Miesowicz nor Rayski) blamed Heisenberg for the behavior of the guards who prevented the Poles from attending his lecture.

I believe it also is unfair to list all the well-known crimes of Hans Frank and the German occupation authorities in Poland and then "return to Heisenberg," suggesting that Heisenberg approved of them or had anything to do with them. Bernstein says that Heisenberg could have refused Frank's invitation. But he does not mention how dangerous that would have been. Heisenberg had postponed his visit to Cracow several times, as one can learn from the papers in Heisenberg's archive, but finally he had to go lest Dr. Frank became hostile to him. Apparently Bernstein does not see the threat in Coblitz' remark that Frank was interested in Heisenberg's visit and would attend his lecture. Dr. Bernstein has never lived under a dictatorship. If Stalin or Hitler or even a SS general like Frank says he is interested in your visit, you may risk your life if you do not show up. Of course, 22 years after the visit Heisenberg may have remembered a letter written by an unknown Cracow institute director on behalf of Frank as a letter written by Frank himself. I don't think that detail is important.

 Heisenberg had been interrogated sharply in the Gestapo cellars some years before. Bernstein mentions the intervention by Heinrich Himmler "which sorted the matter out," but he does not mention that Himmler's mother was involved in this "sorting out" at the initiative of Heisenberg's mother. The two old ladies knew each other well because Himmler's father and Heisenberg's grandfather had been colleagues as directors of Munich high schools. Heisenberg himself was not at all a friend of Himmler, as Bernstein's sentence might suggest. He often dreamt, as he once told me, that he was going to be hanged by the Nazis, as was indeed the fate of several of Heisenberg's friends who were involved in the failed plot of July 20, 1944 against Hitler.

Bernstein mentions that Heisenberg had a report from a colleague at his Institute in Leipzig about the first mass executions of Jews in Poland. This colleague could have been Edwin Gora. Edwin Gora was a Polish citizen of German descent who was studying theoretical physics at Warsaw University when Germany invaded Poland. Warned of imminent arrest, he returned to his home town in South Poland and wrote to Heisenberg. Heisenberg invited him to Leipzig and helped him to enroll at Leipzig University. In 1941 the Gestapo ordered Heisenberg to bar Gora from his institute because of his "politically doubtful mixed extraction." Thereafter, Heisenberg gave him private lessons at his home and enabled him to pass the doctoral examination.

What did Heisenberg know in 1943 about the massacres of Jews in Poland? I do not know but I consider it likely that he had heard of some outrageous events of that type, as Bernstein has mentioned. This is what Heisenberg's wife says in her book from which Dr. Bernstein gathered the information about the private report Heisenberg had received about some of the first massacres. The full extent of the holocaust, however, was not generally known. The general public in Germany at that time only knew that the Jews were being transported to labor camps and settlements somewhere in the East. Even the Jews in Germany themselves believed this, taking along, with official permission, suitcases with their personal belongings. Occasional rumors of killings could be attributed either to enemy propaganda or, if considered to be true, to unauthorized excesses by some local commanders.

The reaction of Heisenberg's father-in-law, as described by Elisabeth Heisenberg in her book, and quoted by Bernstein, shows how impossible it seemed to the average German at the time to believe such reports on massacres committed by Germans. There were, of course, no reports on such crimes in German newspapers or radio broadcasts. Listening to foreign radio stations had been declared a crime punishable by many years in prison or concentration camp, and in "severe cases" by the death penalty, as was the spreading of rumors that would undermine the morale of the army or the "home front." The massacres were classified top secret. Even Himmler, in his infamous Posen (Poznan) speech to SS generals said that these "heroic" massacres will never be mentioned. Anybody who knew about them was, at the risk of the gravest of consequences, not allowed to talk about them. Under these circumstances, it is highly unlikely that Heisenberg was told about the liquidation of the ghettos of Warsaw and Cracow before or during his visit.

Why didn't Heisenberg write about his Cracow visit? Probably he did not do so because the visit had no particular purpose and no particular consequences. It had not happened at his own initiative. Whereas he had gone to Copenhagen in 1941 for a special purpose to discuss with Bohr the difficult questions arising from the ultimate feasibility of nuclear bombs, there was no program of his own connected with the Cracow visit. Its only purpose was not to irritate a high-ranking member of the Nazi hierarchy who had already invited him several times. Heisenberg did not want to run an unnecessary risk. I don't know why Bernstein thinks that Heisenberg's 1943 Cracow visit is an enigma.

Bernstein mentions again what Heisenberg said in Copenhagen in the cafeteria of Bohr's Institute about occupied Poland. To put that in perspective one has to remember a little bit of history. Heisenberg is reported to have regretted German occupation of Denmark, Norway, Belgium and the Netherlands, but to have expressed the view that the Eastern European countries were known to be unable to rule themselves. It is also reported that the well-known Danish physicist Christian Møller, collaborator of Niels Bohr, replied: "So far we only learned that Germany is unable to rule itself!" This view expressed by Heisenberg on the countries of Eastern Europe had been the general view in Germany, and perhaps elsewhere, for centuries. Since the end of the 18th century and up to 1918, only about two decades before Heisenberg's visit to Copenhagen in 1941, Poland had been divided between Russia, Austria, and Prussia. The Baltic states had been part of the empire of the Czar. Czechoslovakia, Hungary and parts of Yugoslavia and Rumania belonged to the Habsburg empire. Before 1795 the position of the Polish king was very weak, and so was the Polish parliament. Any nobleman could veto its decisions. Between the wars, in the 1920s and 1930s, Poland was governed by the semi-dictatorial regime of Pilsudski, and Hungary by that of Admiral Horthy. Yugoslavia and Rumania did not have democratic governments either. When, in July 1915, during the general discussion of German war aims, 191 liberal and moderate German scientists and scholars, among them Max Planck and Albert Einstein, signed a petition against German annexations in the West, arguing that the incorporation or affiliation of politically independent populations or of populations used to independence was to be rejected, they left open the road to territorial expansion in the East.[1] Thus, Heisenberg's remark had nothing to do with approval of Hitler's aggressive policies. It was just a historical reminder based on a view that had been generally held, at least in Germany, for a very long time.[2]

Bernstein says that during a brief period Heisenberg was himself under suspicion. In fact, Heisenberg was under suspicion by some quarters of the Nazi power structure all the time. Already in 1933 he had refused to sign a manifesto drafted by Nazi scientists in support of Hitler. In the following years he was constantly attacked as an adherent of "Jewish physics," culminating in the 1937 article in the SS journal calling him a White Jew. Even though his enemies in the Nazi party were silenced to some extent by Himmler's 1938 judgment that Heisenberg was an unpolitical scientist who should not be killed, his ideological opponents in the Nazi party had not forgotten him and were watching him closely. The suggestion that a visit to Cracow by Heisenberg with his busy schedule was not attractive at all sounds very plausible. He had no mission there as he had had when he visited Bohr in Copenhagen. He accepted reluctantly because refusing could have meant making another enemy in high-ranking circles. Frank was a SS Gruppenführer (General). Living in a criminal dictatorship as a well-known person Heisenberg was used to some extent to being confronted with criminals in high positions and having to cope with them. Admittedly, Heisenberg was not a Nazi. Why should he have enjoyed going to Cracow? Of course, once he had accepted, he also had to accept spending the night in Frank's castle among whatever furniture or art treasures there were. Why should we assume that Heisenberg liked that? Or could he have protested against what he saw? That would have been contrary to the intention of not making enemies. We might only ask, as Mrs. Heisenberg seems to have done, if with more courage Heisenberg could have declined the invitation altogether. But who are we, living in safe democracies, to demand more courage when one more influential enemy could have meant, in the final analysis, a grave risk for himself, his family, and the teaching of his physics?

---

[1] See, for example, Süddeutsche Zeitung of 20/21 July, 2002.
[2] In order to avoid misunderstandings, let me stress that these distorted views are no longer held in Germany. They are a matter of the past. Poland is now a respected ally of Germany, and many Germans, including myself, have close friends in Poland.

Bernstein says that he does not know why Heisenberg was willing to go to Cracow in 1941 when he did not get clearance. It seems to me that it was for the same reason he did go in 1943: Frank had invited him, and he did not want to irritate him.

I give one small comment regarding Bernstein's quotation that Heisenberg was finally allowed by Himmler to teach modern physics as long as he did not mention Jewish authors. I think this quote is correct. However, I own a 1944 edition of Heisenberg's book, "Die physikalischen Prinzipien der Quantentheorie." The index shows the names of Einstein (mentioned 5 times in the book), Born (4 times), Franck, Pauli, and Wigner, all of them Jews by Nazi definition. It seems that Heisenberg managed to disregard Himmler's injunction.

A few additional points: Dr. Bernstein claims that Werner Heisenberg and Hans Frank joined the Neupfadfinder (new pathfinders) about the same time. According to Dr. Helmut Rechenberg, keeper of the Heisenberg archives, there is no evidence from the existing rosters of the Neupfadfinder that Hans Frank was a member at any time.[3] It is not correct that the familiar "Du" used lifelong between former students of the same highschool means "that the friendship was close enough," as Dr. Bernstein seems to suggest. There is no indication in the Heisenberg archives that Heisenberg had any contact with Frank since he left highschool and before he received Frank's invitation to lecture in Cracow. Finally, Dr. Bernstein's English translation of Heisenberg's statement to David Irving about Frank, after losing sight of him, "O.K., I will have nothing further to do with him," is wrong. The correct translation is: "It is a good thing that I have nothing to do with him"[4] It is a sigh of relief whereas the "O.K." of Dr. Bernstein could mean that Heisenberg resigned with regret to the fact that he lost contact with Frank.

Dr. Klaus Gottstein
Emeritus Professor
Max Planck Institute for Physics
Föhringer Ring 6
80805 Munich, Germany

e-mail: Klaus.Gottstein@unibw-muenchen.de

---

[3] Private communication
[4] „gut, dass ich nichts mit ihm zu tun habe."